# Cavity-Enhanced Collective Quantum Processing with Polarization-Encoded Qubits

Kamil Wereszczyński[1][0000-0003-1686-472X], Józef Cyran[1][0009-0006-5205-8986], Adam Brzezowski[1][0009-0004-6997-445X], Dawid Załużny[1][0009-0003-5106-0855], Robert Potoniec[1][0009-0005-7477-3625] Kasper Wiśniowski[1][0009-0004-6696-9778] and Agnieszka Michalczuk[1][0000-0002-8963-1030]

[1] merQlab, Department of Computer Graphics, Vision and Digital Systems, Silesian University of Technology, Gliwice, Poland
kamil.wereszczynski@polsl.pl

**Abstract.** We introduce a cavity-enhanced optical architecture for collective quantum processing in which logical qubits are encoded in the polarization subspace of recirculating intracavity modes. The physical carrier and computational degree of freedom are explicitly separated: harmonic cavity bundles provide a stable resonant substrate, while programmable polarization transformations implement single-qubit operations. A polarization-selective nonlinear interaction in the entanglement region generates tunable controlled-phase gates, enabling a universal gate set. A parameter-scaling analysis shows that order-unity conditional phases are attainable in centimeter-scale cavities using experimentally accessible solid-state nonlinear media, without requiring extreme nonlinear coefficients, millisecond photon lifetimes, or sub-hertz laser stabilization. The results indicate that resonant recirculation provides a physically plausible platform for cavity-based collective quantum architectures.

**Keywords:** Quantum Computing, Qubit, Photonics, Optical Resonator, Optical Cavity.

## 1 Introduction

In the quantum circuit model, algorithms are implemented as sequences of unitary operations acting on qubits. In practice, the dominant physical constraint is often circuit depth: the number of consecutive operations that must be completed within the coherence time of the system. In many photonic implementations, interactions are effectively single-pass: once a quantum state traverses an optical element, it leaves the interaction region. Increasing circuit depth therefore requires cascading additional components or complex re-injection schemes, leading to growing interferometric instability and demanding phase control.

Recirculating and time-multiplexed architectures partially alleviate this limitation by reusing optical paths. However, they typically operate within a fixed logical encoding and do not exploit the intrinsic spectral structure of the cavity as an active computational resource. This motivates the following question: can a cavity be engineered such that the same operators act repeatedly on the same quantum wavepacket within a fixed physical volume, without external re-injection, while preserving controllability and interaction selectivity?

We address this question by proposing a cavity-enhanced architecture in which the harmonic eigenmode structure of an optical resonator serves as a reusable computational substrate. A multimode harmonic bundle confined in each cavity arm provides a stable physical carrier, while logical qubits are encoded exclusively in the polarization subspace of the intracavity field. Local SU(2) operations are implemented via programmable phase shifts and polarization mixing. Entanglement arises from polarization-selective nonlinear interactions in a central entanglement region, leading to effective cross-phase Hamiltonians of the form $\hat{N}_i\hat{N}_j$. The resulting controlled-phase gates, combined with arbitrary single-qubit rotations, form a universal gate set within the polarization Hilbert space.

Quantum computing [1] platforms are currently dominated by superconducting circuits, trapped ions, neutral atoms, and photonic architectures, each offering distinct trade-offs in scalability, coherence, and interaction mechanisms. Superconducting quantum processors use Josephson-junction–based nonlinear oscillators (e.g., transmons) to realize fast, chip-integrated qubits with high-fidelity single- and two-qubit gates [2]. Systems exceeding tens to hundreds of qubits have been demonstrated [3], [4], although scalability is constrained by cryogenic requirements, coherence limits, and connectivity complexity [5].

Trapped-ion quantum computers [6] encode qubits in long-lived internal electronic states of ions confined in electromagnetic traps, with entangling gates mediated by shared motional modes. They have demonstrated record-high single- and two-qubit fidelities and full algorithmic implementations on small- to mid-scale systems [7]. However, scalability is challenged by laser control complexity, motional mode crowding, and architectural constraints in large ion chains or modular trap networks [8].

Photonic quantum computing [9] comprises several distinct paradigms that differ in encoding strategy and the mechanism used to generate effective nonlinear interactions. Linear-optical approaches encode qubits in single photons and implement entangling operations via measurement-induced nonlinearities [10]. While experimentally mature and compatible with integrated photonics, such schemes typically rely on probabilistic gates and significant resource overhead. In contrast, the present cavity-based architecture seeks to provide a deterministic conditional phase through resonant accumulation of weak nonlinearities, potentially reducing reliance on measurement-induced interactions. Weak optical nonlinearities, including cross-Kerr interactions, have been widely studied [11] for deterministic photonic gates but are limited by the very small phase shift per interaction .

Continuous-variable (CV) platforms encode information in optical quadratures and employ Gaussian operations supplemented by non-Gaussian resources such as squeezed states [12], [13]. These systems enable deterministic transformations within the Gaussian regime but require additional nonlinear elements to achieve universality. The proposed model similarly operates in a field-based regime but introduces polarization-selective cross-phase interactions as an intrinsic entangling mechanism, offering an alternative route to non-Clifford operations without relying on extreme squeezing levels. Resonant and cavity-enhanced schemes have also been explored as a mechanism for amplifying weak optical nonlinearities through repeated interaction, providing an alternative to single-pass strong-coupling approaches.

Measurement-based photonic architectures generate large cluster states through spatial or temporal multiplexing and perform computation via adaptive measurements [14], [15]. Although highly scalable in principle, they demand complex state preparation and synchronization. The cavity-enhanced approach explored here instead emphasizes repeated controlled interactions within a fixed physical volume, which may provide a complementary strategy in which entanglement is generated dynamically rather than precompiled into large resource states.

Finally, cavity- and strong-coupling-based photonic systems aim to induce photon–photon interactions using nonlinear media or emitter-assisted coupling [16], [17]. Such approaches often operate at the single-photon level and require strong coupling or high-finesse conditions. In contrast, the present framework considers a collective-field regime in which weak nonlinearities are coherently accumulated through resonant recirculation, potentially relaxing the requirement for extreme single-photon coupling strengths while maintaining operator-level quantum evolution.

In contrast to single-pass photonic schemes, the proposed architecture enables repeated coherent interactions within a fixed physical volume, allowing circuit depth to be effectively traded for interaction time. Importantly, the model is accompanied by an explicit parameter-scaling analysis (Sec. 4), demonstrating that order-unity conditional phases can be achieved within experimentally accessible regimes.

## 2 Physical background

In this section we describe the physical background of the proposed framework. We present the quantum definition of the bundle of harmonic in the optical cavity in macro scale. Furthermore, we define the wave function within polarization component to be able to construct Hamiltonian and justify non-Clifford nature of proposed system.

### 2.1 The architecture of the proposed system

We propose the novel architecture of photonic quantum computation framework consisting of multiple optical macro-cavities arranged in a star-shaped geometry around a common central interaction region, referred to as the Entanglement Area (EA), as illustrated in **Fig. 1**. Each cavity is formed by a pair of coupling mirrors (input and output) and supports a discrete set of harmonic modes that constitute the physical carriers of the logical encoding defined in subsequent sections.

Two types of optical fields are coupled into each cavity: a computational field, responsible for establishing and maintaining the multimode standing-wave structure, and an addressing field, directed toward the EA to enable controlled inter-mode interactions. The central EA serves as a spatially localized nonlinear interaction region in which selected harmonic modes originating from different cavities can interact in a controlled manner.

Within each cavity, local phase and polarization control elements implement single-qubit operations inside the encoded subspace, while interactions mediated through the EA enable multi-qubit coupling. After the computational stage, selected modes are extracted via output couplers and directed to a polarization-resolved detection system, allowing measurement of the encoded logical states.

This architecture separates local unitary control from nonlinear inter-cavity coupling, while maintaining spatial confinement of the optical field within a fixed physical volume. The formal structure of the harmonic-mode Hilbert space and the associated logical encoding are introduced in the following section.

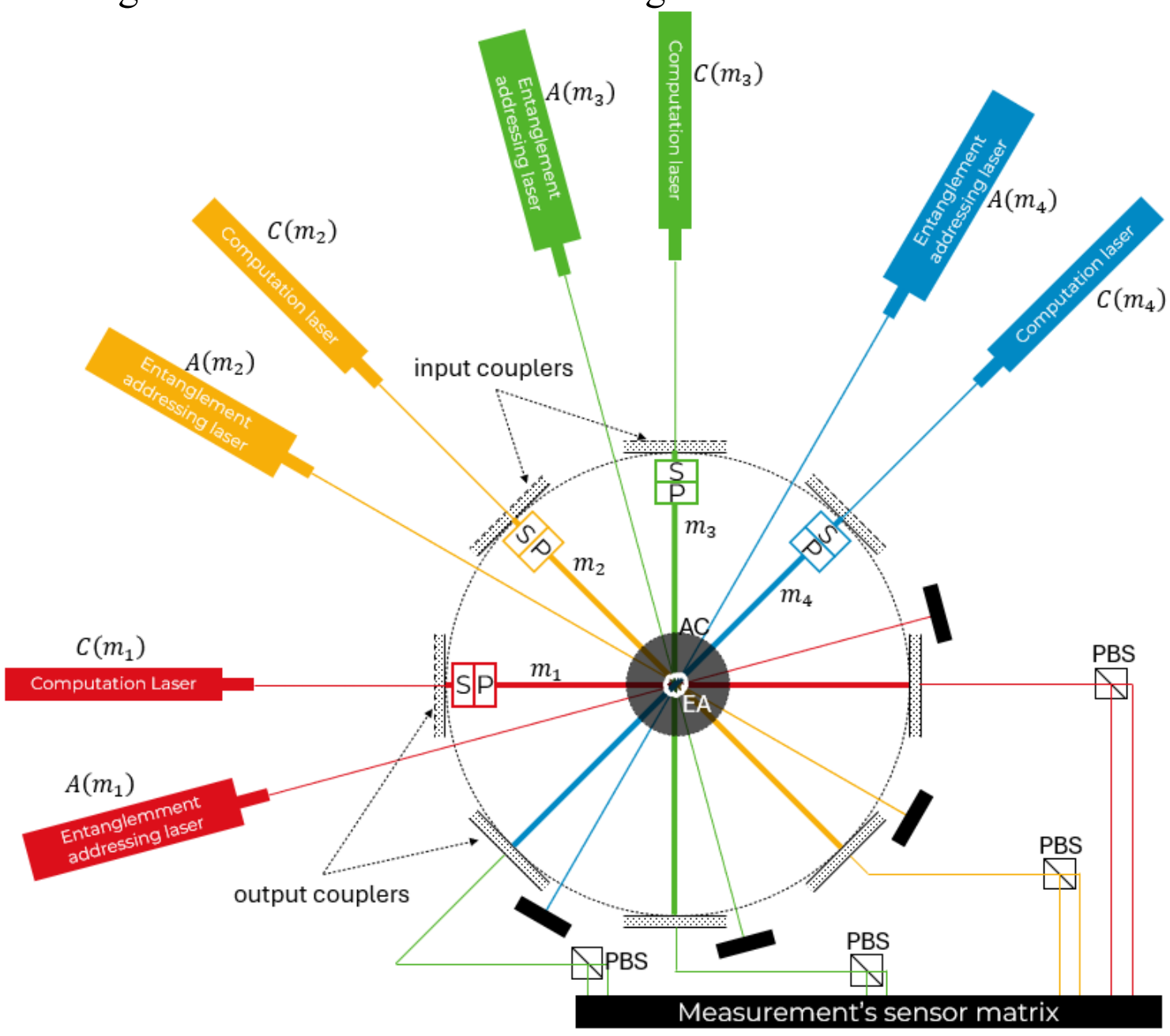

**Fig. 1.** Conceptual architecture of the proposed framework, showing multimode cavities ($m_1 - m_4$), computation and entanglement addressing fields, input/output couplers, and a nonlinear interaction region (EA).

### 2.2 Harmonic Structure of the Cavity Hilbert Space

In free space, the electromagnetic field admits a continuous spectral decomposition into independent modes, each of which can be quantized as a harmonic oscillator, as follows:

$$\hat{E}_c(x,t) = \int d\omega[a(\omega)u_\omega(x)e^{-i\omega t} + h.c.], \quad (1)$$

where $a(\omega)$ and $a^\dagger(\omega)$ are annihilation and creation operators associated with the mode of frequency $\omega$, fulfilling the commutation relation

$$[a(\omega), a^\dagger(\omega')] = \delta(\omega - \omega'). \quad (2)$$

The function $u_\omega(x)$ defines the spatial mode profile determined by the Maxwell equations, while the exponential factor represents its temporal evolution. The equation above describes a field as the superposition of a continuum of independent harmonic oscillators.

In a cavity, boundary conditions discretize the spectrum, replacing the continuous integral by a discrete sum over longitudinal modes. The boundary conditions are expressed as: $L = q\frac{\lambda}{2}$ where $L$ is the length of the cavity, $q \in \mathbb{Z}_+$ is a longitudinal mode index. The condition leads to a discrete set of allowed frequencies $\omega_k$. Consequently, the continuous integral is replaced by a discrete sum over cavity modes, and the field operator takes the discrete form

$$\hat{E}(x,t) = \sum_k (a_k u_k(x) e^{-\mathrm{i}\omega_k t} + a_k^\dagger u_k^*(x) e^{\mathrm{i}\omega_k t}). \tag{3}$$

Here, $a_k$ and $a_k^\dagger$ denote annihilation and creation operators for $k$-th harmonic mode of the cavity and $\omega_k$ is its frequency. Each mode $k$ represents an independent quantum harmonic oscillator described by its own Hilbert space $\mathcal{H}_k = span\{|n_k\rangle : n_k = 0,1,2,\dots\}$, ($|n_k\rangle$ is a number state). Hence, the total cavity Hilbert space is given by the tensor product of these spaces as follows: $\mathcal{H} = \otimes_k \mathcal{H}_k$.

In practice, the driving field has a finite spectral bandwidth. Therefore, only a finite subset of modes carries non-vacuum occupation. We denote by $\mathcal{B} \subset \{k\}$ the selected harmonic bundle used for logical encoding. The corresponding subspace is then given by $\mathcal{H}_\mathcal{B} = \otimes_{k\in\mathcal{B}} \mathcal{H}_k$. The selection of $\mathcal{B}$ is not solely determined by the spectral bandwidth, but also by the logical structure imposed on the computational subspace.

### 2.3 Effective Nonlinear Coupling Between Harmonic Bundles

In the proposed architecture, each arm of the star-shaped configuration supports a harmonic bundle $\mathcal{B}_j \subset \{k\}$, defined in Section 2.2 pg. 4 as a finite subset of cavity modes confined within a given spectral bandwidth. Physically, a bundle corresponds to a multimode optical field propagating within a single cavity arm and repeatedly interacting each other with the EA.

Rather than treating individual photons or single modes as independent carriers of information, we consider the entire bundle as a collective physical entity. The total photon number operator associated with the $j$-th bundle is defined as:

$$N_j = \sum_{k\in\mathcal{B}_j} a_k^\dagger a_k \tag{4}$$

where $a_k^\dagger$ and $a_k$ are the creation and annihilation operators mentioned above. Due to commutation relations, operator $N_j$ describes the total occupation configuration of the multimode field existing in the $j$-th cavity arm of the star-shape system.

Including of the polarization results that each mode $k$ admits two orthogonal components, and the corresponding collective occupation operators may be defined analogously. At this stage, however, we restrict attention to the total bundle occupation as the primary dynamical quantity governing inter-bundle interactions.

In anisotropic solid-state nonlinear media, the third-order susceptibility is tensorial and may exhibit polarization-dependent components determined by crystal symmetry and phase-matching conditions. By appropriate choice of material orientation and operating polarization, the effective nonlinear response can be made to predominantly couple selected polarization modes. Under such conditions, the interaction within the computational subspace acquires an effective form proportional to the product of photon-number operators associated with that polarization component.

**Hamiltonian in Entanglement Area.**

When two or more cavity arms intersect within the Entanglement Area (EA), their corresponding electromagnetic fields spatially overlap inside a common nonlinear medium. In a case, the macroscopic polarization of the medium acquires higher-order contributions with respect to the total electric field. To lowest non-vanishing order beyond the linear response, the nonlinear polarization contains terms proportional to the third power of the field amplitude. The refractive index experienced by one bundle becomes dependent on the intensity of another, consequently.

At the effective level, this intensity-dependent refractive index leads to a phase shift accumulated by the field during propagation through the EA region. Since the phase shift is proportional to the product of field intensities, and intensities are directly related to photon number operators, the corresponding effective interaction Hamiltonian between two bundles $i$ and $j$ may be written in the form

$$H_{ij} = \hbar\chi_{ij}N_iN_j, \tag{5}$$

where $N_i, N_j$ are defined in eq. (4) and $\chi_{ij}$ is an effective nonlinear coupling coefficient determined by the properties of the medium, spatial mode overlap, and optical intensity in the EA. This interaction corresponds to an effective cross-Kerr-type nonlinearity widely considered in photonic quantum computing, here realized through resonantly enhanced multimode field recirculation.

This form represents the lowest-order photon-number-conserving nonlinear interaction compatible with the symmetry of the system. In particular, it preserves the total photon number within each bundle while introducing a conditional phase evolution that depends on the occupation of the interacting arms. The nonlinear coupling can be engineered to act selectively on chosen polarization components, resulting in an interaction term proportional to the product of photon-number operators associated with a specific polarization mode.

The unitary evolution generated by eq. (5) over an interaction time $t$ is

$$U_{ij}(t) = e^{-i\chi_{ij}tN_jN_j}. \tag{6}$$

The essential feature of this dynamics is that the accumulated phase grows linearly with interaction time. Even if the microscopic nonlinear response of the medium is weak, the resonant recirculation of the optical field within each cavity arm allows repeated traversal of the EA region. Since the phase shift per traversal is additive, repeated coherent circulation results in a linear accumulation of the nonlinear phase provided that losses and saturation effects remain negligible over the interaction time. If a single pass through the EA produces a small phase shift $\phi_0$, then after $K$ round-trips the effective phase becomes:

$$\phi_{eff} = K\phi_0. \tag{7}$$

Therefore, the macro-cavity architecture provides a mechanism for enhancing weak third-order nonlinear effects through coherent accumulation rather than through extreme field intensities alone.

For multiple arms intersecting in the EA, the effective interaction generalizes naturally to a sum of pairwise terms,

$$H_{EA} = \hbar \sum_{i<j} \chi_{ij} N_i N_j, \tag{8}$$

which reflects the fact that all overlapping bundles may contribute to the nonlinear polarization of the medium. In the symmetric configuration, and for homogeneous spatial overlap, the coefficients $\chi_{ij}$ may be approximated as equal. In more general situations, their values can be modified by adjusting the local optical intensity or the addressing field associated with selected arms, thereby allowing controlled variation of the effective coupling strength. The feasibility of achieving order-unity conditional phase shifts under realistic experimental conditions is analyzed in Sec. 3.1, pg. 8

**Measurement aspects.** Since the computational degree of freedom is encoded in the collective polarization state of the intracavity field rather than in individual photons, the readout can be performed via intensity measurements after polarization-resolved splitting. In this regime, expectation values of photon number operators correspond directly to measured intensities, allowing the use of high-sensitivity imaging detectors such as EMCCD cameras. Namely, polarization-resolved readout corresponds to measurement of the Hermitian Stokes operator for the $i$-the arm of the star-shaped system:

$$\hat{S}_{i,z} = \hat{N}_{i,H} - \hat{N}_{i,V}. \tag{9}$$

$\hat{N}_{i,H}$, $\hat{N}_{i,V}$ are the photon number operators for the horizontal and vertical polarization modes. The expectation value $\langle \hat{S}_{i,z} \rangle$ is obtained directly from intensity differences after polarization beam splitting. Hence, it links the operator formalism to the measurement method mentioned above.

## 3 The framework for quantum calculations based on harmonic structure in cavity

In this section, we formalize the computational model implemented by the macro-resonator architecture. Although the physical system consists of harmonic bundles of standing-wave modes described in Sec. 2, pg. 3, the computational degree of freedom is encoded exclusively in the polarization subspace of each cavity arm.

Let $i \in \{1, 2, \ldots, M\}$ be indexes of resonator arms like in the illustration **Fig. 1** that intersects in EA. After initialization, in each arm there exists a stabilized harmonic bundle acting as a collective optical carrier. The bundle structure determines the spatial and spectral confinement of the field but does not constitute an independent computational register. The logical qubit associated with arm $i$ is defined as the two-dimensional polarization subspace $\mathcal{H}_i = span\{|H\rangle_i, |V\rangle_i\}$, where $|H\rangle_i, |V\rangle_i$ denote orthogonal (i.e., horizontal and vertical) polarization modes of the intracavity field in arm $i$.

We define a computational basis $\{|0\rangle, |1\rangle\}_i$ for $i$-th arm as:

$$|0\rangle \equiv |H\rangle_i, \quad |1\rangle \equiv |V\rangle_i. \tag{10}$$

The full, computational Hilbert space of the device is therefore given by the tensor product $\mathcal{H}_c = \bigotimes_{i=1}^{M} \mathcal{H}_i$. This space has dimension equal to $2^M$. Entanglement between logical qubits arises through nonlinear interactions described in the Sec. 2.3, pg. 5.

It is important to emphasize that the harmonic bundle structure discussed previously provides a physically robust carrier of the polarization state but does not increase the number of logical qubits. The computational degrees of freedom are strictly polarization-based, while the field operators associated with each arm act within the corresponding Fock space of the polarization modes.

### 3.1 Initialization & one-qubit operations

Initialization proceeds in two stages: stabilization of the harmonic bundle in each cavity arm and preparation of a definite polarization state. Attenuation of non-harmonic components yields a spectrally confined intracavity field that serves solely as a physical carrier of the qubit. The logical register is initialized by setting each arm to the horizontal polarization, defining the state: $|\Psi_0\rangle = |0\rangle^{\otimes M}$.
Coherent operation further requires stabilization of relative optical phases, both between polarization components within each arm and across different arms. Active phase control ensures that the initialized register forms a coherent tensor-product state, preserving the unitary character of subsequent multi-qubit interactions.

The logical qubit is represented by two-dimensional polarization subspace in each of resonators arms. The conventional representation of one qubit is usually given by $\alpha|0\rangle + \beta|1\rangle, |\alpha|^2 + |\beta|^2 = 1, \alpha, \beta \in \mathbb{C}$. Operation on the single qubits corresponds to unitary transformations in $SU(2)$ acting on the polarization subspace $\mathcal{H}_i$. They can be understood as rotations on the Bloch Sphere, which is the common and intuitive way for visualizing the qubits. For construction of single-qubit unitary operators, we introduce two elementary rotations: around the polar axis and around the chosen equatorial axis.

The $SU(2)$ group is generated by rotations about two non-parallel axes, e.g., the z-axis and any equatorial axis: $\{R_z(\varphi), R_\perp(\theta,\gamma)\}$, where the second rotation is around the arbitrary chosen equatorial axis; $\gamma$ denotes the angle between the chosen axis and $x$. The rotations are defined as follows:

$$R_z(\varphi) = e^{-\frac{i\varphi}{2}\sigma_z},$$
$$R_\perp(\theta,\gamma) = e^{-\frac{i\theta}{2}(\cos\gamma\sigma_x + \sin\gamma\sigma_y)}, \tag{11}$$

where $\sigma_x, \sigma_y, \sigma_z$ are Pauli matrices.

Those two operations admit a direct physical interpretation in terms of polarization- and phase-resolved operations on the intracavity field. Namely, a rotation about the $z$-axis of the Bloch sphere $R_z(\varphi)$, corresponds to a controlled differential phase shift between the horizontal and vertical polarization modes in a given arm. In the polarization basis, this operation acts, in polarization basis, as:

$$|H\rangle \to e^{-\frac{i\varphi}{2}}, |V\rangle \to e^{\frac{i\varphi}{2}}. \tag{12}$$

The parameter φ defines the relative phase between the horizontal and vertical polarization components and is implemented as a controllable birefringent phase delay. A rotation about an equatorial axis $R_{\perp}(\theta, \gamma)$ corresponds to coherent mixing of the two polarization modes, where γ determines the orientation of the rotation axis (i.e., the $\sigma_x, \sigma_y$ combination) and $\theta$ sets the mixing angle. Operationally, this is realized through polarization-mode mixing combined with controlled phase offsets. Together, those transformations provide complete local control over the polarization qubit in each resonator arm. Since arbitrary rotations about the $z$-axis and an equatorial axis generate the full SU(2) group, any single-qubit unitary operation $U \in SU(2)$ acting on the polarization subspace $\mathcal{H}_i$ can be decomposed into a finite sequence of the elementary transformations introduced above. In particular, any unitary $U$may be written in the Euler form:

$$U(\varphi, \theta, \lambda) = R_z(\varphi) R_{\perp}(\theta, 0) R_z(\lambda). \tag{13}$$

Thus, three independent real parameters, $\varphi, \theta, \lambda$ suffice to describe any single-qubit unitary transformation. In the present architecture, these parameters are implemented through controlled differential phase shifts (realizing $R_z$) and coherent polarization mixing with adjustable retardance (realizing $R_{\perp}$). Consequently, arbitrary local unitaries can be applied to each polarization qubit prior to nonlinear inter-arm interactions. When applied independently to multiple resonator arms, these local unitary operators combine as a tensor product operation, describing the whole system.

Let's consider the practical implementation of the sequence $H\,P(\varphi)\,H$in a single resonator arm. The arm contains a differential phase shifter (S) with one parameter$\varphi$ and a polarization-mixing element (P) with two parameters, $\theta$ and $\gamma$ realizing $R_{\perp}(\theta, \gamma)$, traversed by the field during each transit between the mirrors.

For the first Hadamard operation, the mixing element is set to $P_\theta = \frac{\pi}{2}$ per transit and $P_\gamma = \pi$ (fixed for the whole gate processing time), while the phase shifter is set to $S = 0$. Since the field passes twice through the element during a half round-trip, the cumulative mixing corresponds to an effective angle of $\pi/8$, realizing the Hadamard transformation (up to a global phase).

The phase operation $P(\varphi)$ is implemented by setting the phase shifter to $S = \frac{\varphi}{2}$ per transit, with the mixing element set to zero (both parameters). After two transits, the accumulated differential phase equals $\varphi$, reproducing the diagonal unitary $\text{diag}(1, e^{i\varphi})$.

Finally, the second Hadamard is realized by restoring the mixing element to $P = \frac{\pi}{16}$ per transit and setting $S = 0$.

In practice, switching the optical elements on a single-transit timescale is not feasible. The effective gate operation must therefore accumulate over multiple cavity transits while S and P remain fixed. The per-transit values are accordingly scaled by the number of contributing passes, such that the cumulative action reproduces the target unitary. The operating regime is chosen as a compromise between element response time and the accumulation of calibration errors over repeated transits.

### 3.2 Controlled Gates

The entangling operation between two resonator arms $i$ and $j$ follows directly from the effective nonlinear Hamiltonian introduced in Sec.2.3, eq. (8). The unitary evolution corresponding to the angle of conditional phase shift $\varphi$ dependent on time $\tau$ is, then given by:

$$\hat{U}_{ij}(\tau) = e^{i\varphi \hat{N}_i \hat{N}_j}, \varphi = \chi_{ij}\tau. \quad (14)$$

Since the logical qubit in each arm is encoded in the polarization subspace $\{|\boldsymbol{H}\rangle, |\boldsymbol{V}\rangle\}$, the nonlinear interaction can be engineered to couple predominantly to a selected polarization mode. In such a configuration, the effective Hamiltonian becomes proportional to the product of photon-number operators associated with that mode, e.g. $\hat{\boldsymbol{N}}_{i,V}, \hat{\boldsymbol{N}}_{j,V}$. The interaction therefore generates an additional phase only when both arms simultaneously occupy the $|\mathbf{1}\rangle$ state. In the computational basis $\{|00\rangle, |01\rangle, |10\rangle, |11\rangle\}$ the resulting two qubit unitary operator takes the form:

$$U_{CP}(\varphi) = diag(1,1,1,e^{i\varphi}), \quad (15)$$

which is the standard controlled-phase gate. The phase parameter $\phi$is determined by the effective interaction strength and the duration (or equivalently, the number of contributing cavity transits) during which the nonlinear coupling in the Entanglement Area is active.

Thus, the architecture naturally realizes a tunable controlled phase operation, with $\varphi$ serving as a programmable two qubit parameter. In particular, choosing $\varphi = \pi$ yields the controlled Z gate, which together with local single qubit gates forms a universal gate set.

The CNOT gate is implemented as a controlled-phase operation supplemented by single-qubit Hadamard transformations on the target arm. First, this operation is applied on target arm (see previous section). The control arm remains unchanged ($S = 0, P = 0$). Next, the addressing lasers of both arms are activated for a controlled interaction time $\tau$, chosen such that the polarization-selective nonlinear coupling in the EA generates an effective phase $\varphi = \pi$. Due to the engineered coupling proportional to $\hat{N}_{V,i}\hat{N}_{V,j}$, this phase is acquired only when both qubits occupy the logical $|1\rangle$ state, implementing a controlled Z operation. Finally, the Hadamard operation is applied again to the target arm, transforming the controlled Z into a CNOT gate.

Since the interaction phase φ is continuously tunable, the architecture supports non-Clifford operations, ensuring computational universality.

# 4 Feasibility and Parameter Regime

The theoretical framework developed in the previous sections establishes the logical and physical structure of the proposed architecture. We now provide a quantitative feasibility analysis of the proposed architecture, explicitly evaluating whether the required conditional phase shifts can be achieved within experimentally accessible parameter regimes. Rather than focusing on a single numerical example, we formulate a systematic parameter-evaluation procedure that allows feasibility to be assessed in a transparent and reproducible manner.

## 4.1 Methodology for Feasibility Assessment

The objective is to determine whether the accumulated nonlinear phase shift $\varphi_{tot}$ can reach values of order $\pi$ under realistic cavity and material parameters. The evaluation proceeds according to the following parameter mapping.

**Table 1.** Input parameters for feasibility assessment

| | | | | | |
|---|---|---|---|---|---|
| Optical wave-length | $\lambda$ | Cavity length | $L_{cav}$ | Nonlinear interaction length | $L_{nl}$ |
| Nonlinear refractive index | $n_2$ | Optical beam waist | $w$ | Continuous-wave optical power | $P$ |
| Cavity quality factor | $Q$ | | | | |

**Table 2.** Derived quantities for feasibility assessment

| | | | |
|---|---|---|---|
| The effective mode area | $A = \pi w^2$ | The intracavity intensity | $I = P/A$ |
| The optical angular frequency | $\omega = \frac{2\pi c}{\lambda}$ | The photon lifetime associated with the cavity quality factor | $\tau = \frac{Q}{\omega}$ |
| The number of effective round trips during the photon lifetime | | | $N_{rt} = \frac{c\tau}{2L_{cav}}$ |

In the light of above, the nonlinear phase accumulated during a single traversal of the interaction region is:

$$\varphi_0 = \frac{2\pi}{\lambda} n_2 I L_{nl}. \tag{16}$$

Furthermore, resonant recirculation leads to coherent accumulation over multiple passes. The total conditional phase is therefore:

$$\varphi_{tot} = \varphi_0 N_{rt} = \frac{2\pi}{\lambda} n_2 I L_{nl} \cdot \frac{c\tau}{2L_{cav}}. \tag{17}$$

The architecture supports a controlled-phase gate when

$$\varphi_{tot} \geq \pi. \tag{18}$$

This condition defines a feasible parameter region in the multidimensional space $(n_2, P, w, Q, L_{cav}, L_{nl})$. The methodology above makes explicit the scaling relations governing the system and allows systematic exploration of conservative, moderate, and aggressive parameter regimes without reliance on a single optimized configuration.

The coherent accumulation assumption requires that the laser coherence time exceeds the cavity photon lifetime:

$$\tau_{coh} = \frac{1}{\Delta\nu} \gg \tau. \tag{19}$$

### 4.2 Parameter Regimes and Scaling Analysis

Using the methodology introduced in previous section, we evaluate three representative operating regimes spanning conservative to ambitious parameter sets in order to assess practical feasibility. The **conservative regime** assumes lower-end nonlinear coefficients, moderate optical power and confinement, and cavity quality factors consistent with present high-finesse laboratory systems. The **moderate regime** adopts intermediate nonlinear response and improved confinement with elevated yet realistic power and Q values. The **aggressive regime** assumes upper-range nonlinear coefficients, strong confinement, high circulating power, and very high cavity quality factors, remaining within reported experimental limits.

**Table 3.** Representative input parameter sets

| Variant | $n_2\left[\frac{m^2}{W}\right]$ | $P\,[W]$ | $w\,[\mu m]$ | $Q$ | $\tau\,[\mu s]$ |
|---|---|---|---|---|---|
| **Conservative** | $1\cdot10^{-18}$ | 20 | 30 | $5\cdot10^{9}$ | 2.6 |
| **Moderate** | $5\cdot10^{-18}$ | 30 | 25 | $7\cdot10^{9}$ | 3.6 |
| **Aggressive** | $1\cdot10^{-17}$ | 40 | 20 | $1\cdot10^{10}$ | 5.2 |

For each regime, the total accumulated nonlinear phase $\varphi_{\text{tot}}$ is computed according to the scaling relation derived in previous section. In addition, the corresponding photon lifetime $\tau$ is obtained from the assumed cavity quality factor. The coherence requirement for executing multiple sequential gate operations is expressed through the condition:

$$\Delta\nu = \frac{1}{N\tau}, \tag{20}$$

where $\Delta\nu$ is the laser linewidth and $N$ is the number of coherent gate operations.

| Variant | $\varphi_{tot}\,[rad]$ | $\Delta\nu_{max}$ | | |
|---|---|---|---|---|
| | | **10 ops** | **100 ops** | **1000 ops** |
| **Conservative** | **1.2** | 38 kHz | 3.8 kHz | 380 Hz |
| **Moderate** | **4.1** | 27 kHz | 2.7 kHz | 270 Hz |
| **Aggressive** | **5.2** | 19 kHz | 1.9 kHz | 190 Hz |

The analysis indicates that order-unity nonlinear phase shifts are achievable without millisecond-scale photon lifetimes or extreme nonlinear coefficients. In the moderate and aggressive regimes, $\varphi_{\text{tot}}$ exceeds $\pi$, supporting direct implementation of controlled-phase gates. Even in the conservative regime, the accumulated phase approaches unity, indicating proximity to the operational threshold.

The required laser linewidth for coherent execution of tens to hundreds of operations lies in the kilohertz range, substantially above sub-hertz stabilization levels. Consequently, cavity quality factor and nonlinear material properties constitute the dominant design constraints, rather than extreme laser coherence.

In practical implementations, optical losses (scattering and absorption) and phase noise will limit the effective number of coherent round trips, thereby constraining the achievable interaction time without altering the fundamental scaling relations derived above. Overall, the parameter mapping demonstrates that the proposed resonant enhancement mechanism operates within a physically plausible region of experimentally accessible parameters.

### 4.3 Numerical Fidelity Analysis of the Accumulated Gate Implementation

The curves on **Fig. 2** below were obtained by numerically simulating the accumulated unitary evolution corresponding to the HPH sequence implemented through repeated per-transit applications of the polarization mixing and phase-shift operators. At each reflection, a zero-mean Gaussian perturbation with a prescribed standard deviation was added to the nominal rotation angle. For each parameter setting, the resulting output state was compared to the reference state (noise σ=4×10^(-4)) and the gate fidelity was computed. The reported values represent averages over multiple independent stochastic realizations in order to suppress single-trajectory fluctuations.

**Fig. 2.** Average gate fidelity of the accumulated $HPH$ circuit as a function of Gaussian phase noise (left, $N = 3000$) and as a function of the number of reflections (right, $\sigma = 4 \times 10^{-4}$), obtained from Monte Carlo simulations of per-transit perturbations.

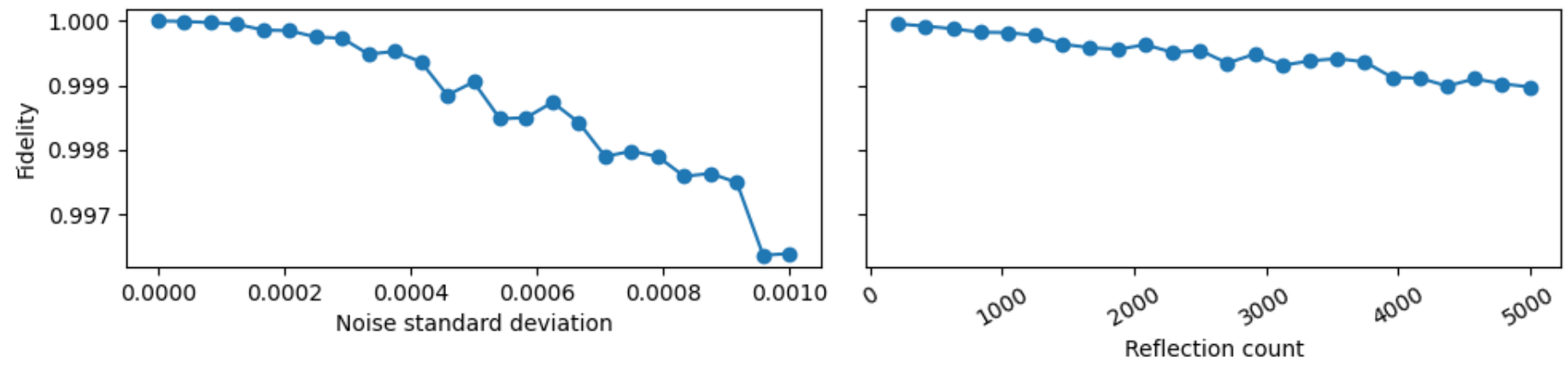

The numerical results demonstrate a controlled and gradual degradation of gate fidelity under cumulative per-transit phase perturbations. For a fixed number of reflections ($N = 3000$), the fidelity decreases smoothly with increasing Gaussian noise standard deviation, exhibiting the expected quadratic sensitivity in the small-noise regime. Even for noise levels approaching $10^{-3}$ rad per transit, the fidelity remains above 0.996, indicating substantial robustness of the accumulated implementation against moderate stochastic fluctuations. Conversely, for a fixed noise level ($\sigma = 4 \times 10^{-4}$), the fidelity shows a slow and approximately monotonic decrease with increasing reflection count, confirming the predicted scaling with the total number of unitary iterations. No abrupt instabilities or chaotic behavior are observed within the investigated parameter range, suggesting that the degradation mechanism is governed by predictable phase diffusion rather than nonlinear amplification effects.

The parameter analysis shows that order-unity nonlinear phase shifts required for controlled operations can be achieved within experimentally realistic cavity and material regimes, with cavity quality factor and nonlinear response emerging as the key design constraints. The accompanying numerical simulations confirm that fidelity degradation under stochastic per-transit noise follows predictable phase-diffusion scaling without instability. Together, the analytical and computational results indicate that the proposed resonant accumulation mechanism operates within a physically feasible and controllable parameter space.

## 5 Discussion and Conclusion

The proposed macro-resonator architecture combines resonant field recirculation with polarization-encoded qubits, separating the harmonic cavity bundle as a physical carrier from the polarization subspace as the computational degree of freedom. This enables repeated coherent interactions within a fixed volume while preserving a clear tensor-product logical structure.

Formally, the system supports universal quantum computation: arbitrary single-qubit SU(2) operations are implemented via polarization mixing and phase shifts, and a polarization-selective nonlinear interaction in the Entanglement Area generates a tunable controlled-phase gate.

Feasibility analysis shows that order-unity nonlinear phases can be achieved in centimeter-scale cavities using accessible solid-state media, with the accumulated phase scaling as $n_2 I\tau$. Required laser coherence lies in the kilohertz regime for tens to hundreds of gates, making cavity losses and nonlinear efficiency the dominant constraints.

The architecture operates in a collective-field regime, where logical qubits are encoded in coherent polarization states of many-photon intracavity modes. Provided phase coherence and low dissipation are maintained, the nonlinear dynamics enable entanglement between resonator arms, with coherent stability representing the primary experimental challenge. properties. Compared to conventional photonic approaches, the proposed scheme shifts the implementation burden from strong single-pass nonlinearities toward coherent accumulation of weak interactions, providing a potentially more scalable route to multi-qubit operations. The framework establishes a consistent theoretical and parametric foundation for further experimental investigation of cavity-enhanced collective quantum architectures.

**Acknowledgements.** The authors would like to acknowledge that this paper has been written based on the results achieved within the OptiQ project. This Project has received funding from the European Union's Horizon Europe programme under the grant agreement No 101080374-OptiQ. Supplementarily, the Project is co-financed from the resources of the Polish Ministry of Science and Higher Education in a frame of programme International Co-financed Projects. Disclaimer Funded by the European Union. Views and opinions expressed are however those of the author(s) only and do not necessarily reflect those of the European Union or the European

Research Executive Agency (REA–granting authority). Neither the European Union nor the granting authority can beheld responsible for them.

This work was funded by the Polish state budget under the program of the Minister of Science and Higher Education "Student Research Clubs Create Innovations", project no. SKN/SP/631635/2025

This preprint has not undergone peer review or any post-submission improvements or corrections. The Version of Record of this contribution is published in LNCS,volume 16789 "Computational Science – ICCS 2026 Workshops", and is available online at https://doi.org/10.1007/978-3-032-29918-5_6